\documentclass[sigconf]{acmart}

\usepackage{multirow}
\usepackage{enumitem}
\usepackage{graphics,balance}
\usepackage{soul}
\usepackage{hyperref}
\usepackage{cleveref}

\definecolor{myBlue}{RGB}{0, 119, 255}
\definecolor{myPink}{RGB}{213, 55, 156}
\definecolor{myGreen}{RGB}{141, 244, 165}
\definecolor{myOrange}{RGB}{245, 146, 110}
\definecolor{revisedColor}{RGB}{255,178,178}

\AtBeginDocument{%
  }

\copyrightyear{2023}
\acmYear{2023}
\setcopyright{acmlicensed}
\acmConference[CHI '23]{Proceedings of the 2023 CHI Conference on Human Factors in Computing Systems}{April 23--28, 2023}{Hamburg, Germany}
\acmBooktitle{Proceedings of the 2023 CHI Conference on Human Factors in Computing Systems (CHI '23), April 23--28, 2023, Hamburg, Germany}
\acmPrice{15.00}
\acmDOI{10.1145/3544548.3580988}
\acmISBN{978-1-4503-9421-5/23/04}

\begin{document}

\newcommand{\classes}{classes} 
\newcommand{\Classes}{Classes}
\newcommand{\class}{class}

\newcommand{\participants}{experts} 
\newcommand{\participant}{expert}

\newcommand{\limitations}{limitations}
\newcommand{\limitation}{limitation}


\newcommand{\revised}[0]{}
\newcommand{\highlight}[0]{}

\title{Memory Manipulations in Extended Reality}

\author{Elise Bonnail}
\affiliation{%
  \institution{LTCI, INFRES, Telecom Paris, IP Paris}
  \city{Palaiseau}
  \country{France}
}
\email{elise.bonnail@telecom-paris.fr}

\author{Wen-Jie Tseng}
\affiliation{%
  \institution{LTCI, INFRES, Telecom Paris, IP Paris}
  \city{Palaiseau}
  \state{}
  \country{France}
}
\email{wen-jie.tseng@telecom-paris.fr}

\author{Mark McGill}
\affiliation{%
  \institution{University of Glasgow}
  \city{Glasgow}
  \state{}
  \country{Scotland, UK}
}
\email{mark.mcgill@glasgow.ac.uk}

\author{Eric Lecolinet}
\affiliation{%
  \institution{LTCI, INFRES, Telecom Paris, IP Paris}
  \city{Palaiseau}
  \state{}
  \country{France}
}
\email{eric.lecolinet@telecom-paris.fr}

\author{Samuel Huron}
\affiliation{%
  \institution{CNRS i3 (UMR 9217), SES, Telecom Paris, IP Paris}
  \city{Palaiseau}
  \state{}
  \country{France}
}
\email{samuel.huron@telecom-paris.fr}

\author{Jan Gugenheimer}
\affiliation{
  \institution{TU Darmstadt}
  \city{Darmstadt}
  \country{Germany}\\
}
\affiliation{%
  \institution{LTCI, INFRES, Telecom Paris, IP Paris}
  \city{Palaiseau}
  \state{}
  \country{France}
}
\email{jan.gugenheimer@tu-darmstadt.de}

\renewcommand{\shortauthors}{Bonnail et al.}

\begin{abstract}
 
Human memory has notable \limitations\ (e.g., forgetting) which have necessitated a variety of memory aids (e.g., calendars). 
As we grow closer to mass adoption of everyday Extended Reality (XR), which is frequently leveraging perceptual \limitations\ (e.g., redirected walking), it becomes pertinent to consider how XR could leverage memory \limitations\ (forgetting, distorting, persistence) to induce memory manipulations.
As memories highly impact our self-perception, social interactions, and behaviors, there is a pressing need to understand XR Memory Manipulations (XRMMs). 
We ran three speculative design workshops (n=12), with XR and memory researchers creating 48 XRMM scenarios. 
Through thematic analysis, we define XRMMs, present a framework of their core components and reveal three \classes\ (\emph{at encoding, pre-retrieval, at retrieval}). Each class differs in terms of technology (AR, VR) and impact on memory (influencing quality of memories, inducing forgetting, distorting memories).
We raise ethical concerns and discuss opportunities of perceptual and memory manipulations in XR. 

\end{abstract}



\maketitle

\section{Introduction}
Human memory is known to have a set of \limitations. Long-term memories can be forgotten (when they are inaccessible), distorted (when they differ from the original perception), or persistent (when they cannot be forgotten) \cite{schacter2002}. As our memories influence our thoughts and behaviors, memory \limitations\ can have a significant detriment to our lives and social relationships. Addressing these \limitations\ is what has given rise to the continual development of new archival mediums (writing, pictures, video) and memory aids (from post-it notes to AI assistants). Seeing the recent rise of Extended Reality (XR) consumer technology (e.g., Meta, Apple, Microsoft), we argue it has the potential to become a ubiquitous technology, able to augment and impact the human intellect, and in particular memory \cite{engelbart1962augmenting}. One example would be the ability to revisit past memories in Virtual Reality (VR). Using immersive 3D reconstructions of past events creates a sense of presence and embodiment, which enhances episodic memory performance \cite{makowski2017, smith2019} and offers a more attractive way to reminisce \cite{xu2020}.

However, while memory \limitations\ can be compensated, they can also be exploited to manipulate memory. Due to the suggestibility of memory, altering a picture can distort memories related to it and lead to the creation of false memories \cite{risko2019, nash2009, wade2002}. XR being a richer media than a regular 2D picture, it could facilitate the creation of false memories \cite{segovia2009}, and more generally facilitate memory manipulations. In addition, XR has demonstrated its ability to take advantage of human perceptual limitations to manipulate users, which we refer to as perceptual manipulations. Much research is focused on Virtual-Physical Perceptual Manipulations (VPPMs) \cite{tseng2022}, which exploit the limitations of human perception in order to effect changes in users’ physical movements (e.g., redirected walking \cite{sun2018}, haptic retargeting \cite{azmandian2016}). Similar to perception, memory has \limitations\ (forgetting, distorting, persistence \cite{schacter2002}) that could be exploited in XR to induce memory manipulations, which we refer to as \textit{XR Memory Manipulations} (XRMMs). Although VPPMs can enhance users’ XR experience (e.g., by enabling a longer virtual walking experience despite limited physical space \cite{sun2018}), these manipulations are generally imperceptible and the intent behind the use of VPPMs is open to abuse \cite{casey2021,tseng2022}. While these perceptual manipulations have an immediate impact on users' physical actions, memory manipulations could be more invasive and have longer-term impact on users' self-perception, social interactions, and behaviors (through the three autobiographical memory functions \cite{bluck2005}). Therefore, it is crucial that we understand the extent to which our memory can be manipulated with XR, and how this technology can uniquely contribute to or amplify these manipulations.

In this paper, we strive to understand what types of memory manipulations are possible in XR, and how the technology works as an enabler or amplifier in these manipulations. To broadly explore the scope of XRMMs, we ran three speculative design workshops, each with two XR researchers and two cognitive psychology researchers having strong knowledge about memory (n=12). We used speculative design methods to envision future scenarios while still grounding the workshop in our current understanding of memory processes and XR capabilities. During these workshops, \participants\ were asked to individually generate scenarios of XRMMs, resulting in 48 XRMM scenarios. The scenarios focused on specific questions: who would use the application, how the manipulation would work and why would it be used, grounded on which cognitive process and XR specific features. Consecutively, \participants\ selected their most promising scenarios that were additionally discussed in more depth to assess, from both technical and cognitive point of views, if they are likely to be implemented in the future, and to discuss their outcomes (22 discussed XRMM scenarios). Three authors coded the data and used thematic analysis in seven sessions (each four hours) to uncover underlying themes and mechanisms used. 

From this analysis, we define XR Memory Manipulations (XRMMs) as \emph{perceptual manipulations grounded in memory limitations with the intention to impact memories, leading to long-term effects on users, through the self, social and directive functions of autobiographic memory}. From the data, we discovered five core components that characterize memory manipulations (\emph{context, memory mechanism, implementation, memory functions, and final goal}). We present these as the framework of XRMMs.
Additionally, we found three \classes\ of XRMMs that differ in terms of technology and outcomes: 1) manipulations occurring \emph{at encoding} (i.e., simultaneously as the original perception of the event), which aim at \emph{influencing the quality of memory encoding} and predominantly \emph{rely on Augmented Reality (AR)} technologies (e.g., steering user attention with AR to make the encoding of memories stronger); 2) manipulations occurring \emph{at retrieval} (i.e., when past events are digitally revisited), which aim at \emph{distorting the perception of a past event} and predominantly \emph{rely on immersive reconstruction of scenes in Virtual Reality (VR)} (e.g., altering a recording of prior event to overwrite the original memory); 3) manipulations occurring \emph{pre-retrieval} (i.e., just before the memory is remembered), which aim at \emph{inducing forgetting} (e.g., hiding elements in AR to help someone to forget something).
\revised{Most of the scenarios take advantage of features unique to XR, which enable new types of memory manipulations. Although some XRMM scenarios do not fully depend on XR to be delivered and are adapted from already existing techniques using more traditional media (e.g., pictures), we present a discussion about how XR-specific features (e.g., the capacity to fully immerse users in a virtual environment) could amplify the effect of memory manipulations.}
While several scenarios have a positive intent (13 positive, 9 negative discussed scenarios), some of them could, intentionally or not, represent harms that need to be addressed. During the discussion, most experts raised ethical concerns and were worried about future abuse. 

We present a final discussion around the potential benefits XRMMs could generate, their potential dangers, and what role our community of HCI researchers is playing in the pursuit of researching and improving perceptual manipulations in XR. We argue that XRMMs are a new type of perceptual manipulation in XR, whose focus is not on impacting immediate actions but on impacting users' on the long-term, through the self, social and directive functions of autobiographical memory.



This work has two main contributions: 1) The definition of XRMMs and a framework presenting the core components that characterize these manipulations; 2) The classification of three \classes\ of memory manipulation in XR and insights about their unique technical characteristics and outcomes (manipulations \emph{at encoding} influencing the quality of memories with AR, manipulations \emph{at retrieval} distorting memories with VR, manipulations \emph{pre-retrieval} inducing forgetting), grounded in ongoing cognitive psychology and XR research.

\section{Background}
\label{sec:bkg}
In this section we present a necessary background about human memory (processes, impacts and limitations) grounded in cognitive psychology research, and our premise of ubiquitous XR technology. The knowledge we present in this section was also explained as an introduction to our workshop, to bridge the gap between experts from both fields.

\subsection{Human Memory}
\label{sec:humanmemory}
Memory refers to the cognitive process of remembering information. It is a broad term, as there are many different types of memory, each with specific capacities and functions. The model proposed by Shiffrin and Atkinson differentiates sensory and short-term memory from long-term memory \cite{shiffrin1969}. In this paper, we focus on long-term memory, which stores information that have been processed thoroughly, as it is the type of memory for which we use technology support more frequently (section~\ref{sec:memoryaids}). The formation of long-term memories occurs in three stages \cite{melton1963, shiffrin1969}: \emph{encoding} (acquiring of new information captured by the senses), \emph{storage} (maintaining and retaining information over time), and \emph{retrieval} (recalling or recognizing previously stored information). 

\subsubsection{Autobiographical Memory Functions}
\label{sec:amfunctions}
In this work, we particularly focus on Autobiographical Memories (AM), which includes recollected memories of personal experiences and general knowledge of the self \cite{eysenck2020}, as it defines and impacts many aspects of an individual's life. Bluck and Alea \cite{bluck2005} name three AM functions: 1) the \emph{self function}, which maintains self coherence over time (e.g., personality, personal beliefs); 2) the \emph{social function}, which helps developing and maintaining connections with others (e.g., remembering conversations with friends or being able to tell an anecdote); 3) the \emph{directive function}, which guides present and future thoughts and behaviors (e.g., deciding to buy something based on previous experience). Unlike other types of perceptual manipulations (e.g., VPPMs \cite{tseng2022}) which have a brief and immediate impact on users, memory manipulations could have a longitudinal and long-term impact on users, through the three AM functions (e.g., impacting their self-perception, social interactions and behaviors).


\subsubsection{Memory Limitations}
\label{sec:limitations}

\begin{table*}[t]
    \centering

    \begin{tabular}{@{}llll@{}}
\toprule
\multicolumn{2}{l}{Memory limitation}                                                                                          & Description                                                                                      & Example                                                                                      \\ \midrule
                                                                                                           & \textbf{Transience}        & \begin{tabular}[c]{@{}l@{}}Decreasing accessibility of\\ memory over time\end{tabular}           & \begin{tabular}[c]{@{}l@{}}Forgetting what exactly \\ happened 10 years ago\end{tabular}     \\
\begin{tabular}[c]{@{}l@{}}\textbf{Forgetting}:\\ When we fail to bring to \\ mind an information\end{tabular}      & \textbf{Absent-mindedness} & \begin{tabular}[c]{@{}l@{}}Inattentive processing contributes \\ to weak memories\end{tabular}   & \begin{tabular}[c]{@{}l@{}}Forgetting location of \\ keys or glasses\end{tabular}            \\
                                                                                                           & \textbf{Blocking}          & \begin{tabular}[c]{@{}l@{}}Temporary inaccessibility\\ of memory\end{tabular}                    & \begin{tabular}[c]{@{}l@{}}Temporarily forgetting a \\ word\end{tabular}                     \\ \midrule
                                                                                                           & \textbf{Misattribution}    & \begin{tabular}[c]{@{}l@{}}Attributing a memory to \\ the wrong source\end{tabular}              & \begin{tabular}[c]{@{}l@{}}Confusion between imagined \\ and real events\end{tabular}        \\
\begin{tabular}[c]{@{}l@{}}\textbf{Distorting}:\\ When a memory differs from\\ the original perception\end{tabular} & \textbf{Suggestibility}    & \begin{tabular}[c]{@{}l@{}}New information can be implanted\\ to a memory\end{tabular}           & \begin{tabular}[c]{@{}l@{}}Inducing false memories \\ through leading questions\end{tabular} \\
                                                                                                           & \textbf{Bias}              & \begin{tabular}[c]{@{}l@{}}Influence of current knowledge and\\ beliefs on memories\end{tabular} & \begin{tabular}[c]{@{}l@{}}Current mood influences \\ perception of the past\end{tabular}    \\ \midrule
\begin{tabular}[c]{@{}l@{}}\textbf{Persistence}:\\ When a memory cannot be\\ forgotten\end{tabular}                 &                   & \begin{tabular}[c]{@{}l@{}}Inability to forget unwanted \\ memories\end{tabular}                 & \begin{tabular}[c]{@{}l@{}}Flashbacks of traumatic \\ memories\end{tabular}                  \\ \bottomrule
\end{tabular}

    \captionof{table}{Seven sins of memory presented by Daniel Schacter \cite{schacter2002}. We refer to them as memory limitations}
    \label{fig:sevensins}
    \Description{Table summerizing the seven memory limitation, classified by Daniel Schacter as "the seven memory sins". The first column presents three high categories (forgetting, distorting and persistence), the second column presents sub-categories (transience, absent mindedness, blocking, misattribution, suggestibility and bias). The third column gives a short description of each limitation, and the last column a simple example of it.}
\end{table*}

Memory plays a ubiquitous role in our lives, but many errors can occur at each stage of the memory process.
Schacter et al. classified memory \limitations\ into seven basic “sins”, that can be separated into three types: \textit{forgetting, distorting, and persistence} (Table~\ref{fig:sevensins}).
We give more explanations about these memory \limitations\ when discussing the grounding of XRMM classes (sections \ref{sec:encoding}, \ref{sec:retrieval}, \ref{sec:preretrieval}).

We argue that these memory \limitations\ are the equivalent of perceptual thresholds leveraged in VPPMs. They are limitations in the cognitive system which applications could leverage or overcome to achieve a desired outcome around the user.

\subsection{Extended Reality (XR)}
\label{sec:xr}
Our research works under the premise that future XR technologies will become ubiquitous (similar to smartphones nowadays), and that users will engage with it on a daily basis, in every life context (mobile, stationary \cite{gugenheimer2016, gugenheimer2019}). We envision a technology that is not focusing on either AR or VR but presents one device (whatever the shape might be) that is capable to function across the whole spectrum of the Milgram Continuum \cite{milgram1994}. This would provide the user the ability to augment the real world (AR), but also the ability to be immersed in a fully virtual scene when necessary (VR). We refer to this concept as ubiquitous XR.

\section{Related Work}

\subsection{Memory Aids}
\label{sec:memoryaids}
HCI research about memory has mainly been focused on memory aids, aiming to compensate memory failures, to improve performance on everyday tasks requiring memory \cite{jamieson2014}. Charness et al. distinguish two types of memory supportive technology: internal memory aids, that support the memory processes, and external memory aids, that substitute and offload memory processes \cite{charness2012}. However, offloading memory can also impair memory and expose users to manipulations \cite{eliseev2021, risko2019}.

\subsubsection{Internal Memory Aids}
\label{sec:internalaids}
Internal memory aids are techniques that support the cognitive process of memorizing, to make it more efficient \cite{charness2012}. One example is the method of Loci, which helps to memorize information by mentally associating them to familiar places \cite{spence1984memory}. HCI research has explored how this memorization technique can be done using visual displays \cite{legge2012, doolani2021, das2019}, and how mnemonics can help users remembering information such as passwords \cite{doolani2021,al-ameen2015,bonneau2014, joudaki2018, castelluccia2017}. The method of Loci has been adapted to XR using virtual memory palaces \cite{vindenes2018, rosello2016, yamada2017,legge2012a}, as immersion can aid recall \cite{krokos2019}. In this paper, we explore in a broader way how XR specific functionalities could be leveraged to influence cognitive processes (e.g., to support or impair memorization).

\subsubsection{External Memory Aids}
\label{sec:externalaids}
External memory aids aim to substitute memory and offload the cognitive process of memorizing (using external storage to offload the quantity of memories stored in mind) \cite{charness2012}. 
These tools can substitute prospective memory (memory of actions to do), to ensure that we remember things we intend to do in the future (e.g., using calendar to remember to attend an appointment). New technologies (e.g., smartphones) provide effective reminder tools (e.g., Google calendar \cite{mcdonald2011}) \cite{jamieson2014, caprani2006, shin2021, jamieson2017}. External memory aids can also substitute retrospective memory (memory of past events), to increase the amount of past events that can be recalled (e.g., by storing pictures in a computer) \cite{jamieson2014, caprani2006}. 
With the ubiquity of smartphones and data storage becoming cheaper, people can record anything at any time, a practice called lifelogging \cite{gurrin2014,harvey2016}. 
These technologies support reminiscence and reflection \cite{cosley2009, isaacs2013}, which improves well-being \cite{bryant2005, isaacs2013}. 
The rich contextual information of images facilitates autobiographical recollection \cite{sellen2007, berry2007}. 
Instead of viewing recordings on a flat screen, these could be stored as 3D scenes that one could revisit with a VR HMD. This idea is started to be explored by companies and artists 
\cite{facebook2018, zotero1915}. 
VR can be used for reminiscence therapy (the discussion of past experiences with other people to improve well-being \cite{woods2018}), by immersing patients in familiar environments in VR \cite{tsao2019, xu2020, siriaraya2014}, and is a more attractive way to reminisce than pictures \cite{xu2020}.

All these prior works demonstrate that technology has the ability to impact users' memory and that HCI is currently heavily engaged in providing tools to enable such. However, our work aims to understand how future memory tools would look like in XR, and provide a critical reflection on potential abuse and opportunities.

\subsection{Risks of Offloading Memory}
Although these memory aids have many benefits, researchers started to question whether such technologies could impair memory \cite{schacter2022, eliseev2021, clinch2021,risko2019} \revised{and pose safety and privacy issues \mbox{\cite{davies2015, bexheti2019a}}}. First, recording tools can divide attention, and affect how events are encoded \cite{eliseev2021,henkel2014}. Additionally, being able to remember everything through a total externalization of memories can have negative effects (e.g, the inability to move on after a breakup), which should be considered in the design of such technologies \cite{sas2013}. Clinch et al. identify two potential risks associated with memory externalization: \emph{inhibition} (retaining less information by ourselves) and \emph{distortion} (errors in what and how we remember) \cite{clinch2021}. Researchers showed that offloading memories to external media exposes them to manipulations (e.g., altering pictures), which can lead to the creation of false memories (i.e., a strong confidence in the memory of an event that did not actually occur) \cite{clinch2021, risko2019, nash2009, wade2002}. 
These questions of memory manipulations are still new and mainly focused on more traditional media like pictures. Our work explores how new immersive technologies like XR could be used to manipulate users’ memory.

\subsection{Perceptual Illusions and Manipulations}
XR devices create visual, auditory, and haptic stimuli, which facilitates control over what users perceive. XR is based on illusions \cite{gonzalez-franco2017} and the creation of immersive virtual environments, in which users have a sense of presence \cite{draper1998, slater1997}. Having a plausible virtual environment was also shown to make users behave realistically (e.g., plausibility illusion \cite{sanchez-vives2005, slater2009}), and feel like their virtual avatar has become a part of their body (e.g., embodiment illusion \cite{spanlang2014}).

These illusions enable the implementation of a variety of perceptual manipulations, such as VPPMs \cite{tseng2022}, which leverage visual dominance \cite{posner1976} and unawareness of sensory discrepancy \cite{vanbeers2002} to influence physical actions. Redirected walking \cite{razzaque2002,steinicke2010,sun2018} and haptic retargeting \cite{azmandian2016, cheng2017} techniques can steer users' physical walking paths and hand movements by imperceptibly manipulating the virtual scene. Mhaidli et Schaub also identified manipulation techniques unique to XR that could be used for advertising purposes \cite{mhaidli2021} (e.g., altering what consumers see to bias their consumption behavior). Although these manipulation techniques could be beneficial for users (e.g., VPPMs enhancing VR experience), researchers started to explore how they could be exploited maliciously to harm users \cite{tseng2022, casey2021,mhaidli2021}, \revised{and how users would react to such attacks \mbox{\cite{cheng2023}}}.

While most of these perceptual manipulations target to have an immediate impact on the user, XRMMs have the potential to have longer-term impacts, through the AM functions (section~\ref{sec:amfunctions}).

\subsection{XR Memory Manipulations}
These XR perceptual manipulations are possible because users’ behaviors in the virtual environment are similar to real behaviors. 
Studies showed that the encoding and retrieval processes of virtual events are similar to those of real events, making VR an ideal tool for studying real-life cognition under controlled conditions \cite{kisker2021, picard2017, lenormand2022}. Particularly, immersion, embodiment, and sense of presence enhance memory encoding and retrieval \cite{smith2019, makowski2017, dinh1999, blanke2018, penaud2022}.

While false memories have traditionally been studied using narratives and pictures, Segovia et al. showed that immersing preschool children in a virtual experience can generate false memories \cite{segovia2009}. As pointed out by Aliman et al., combining the capacity of AI to generate deepfakes with the capacity of VR to create durable memories could be used for malicious creation of false memories \cite{aliman2020}. 

HCI research focused on how to compensate and overcome memory \limitations, and what risks could represent such techniques. However, little work has explored how technologies could leverage memory \limitations\ to achieve other purposes. An example of such work is \textit{Mise Unseen}, which takes advantage of visual memory failures and the change blindness effect to implement a redirection technique in VR \cite{marwecki2019}. Friedman et al. introduced a method to give people the illusion of time travel in VR, which can have an impact on implicit morality and evaluation of past events \cite{friedman2014}, and Cuperus et al. showed that manipulating replays of events in virtual reality can alter the memory for sport performances and impact the feeling of competence \cite{cuperus2016}. While all these prior works demonstrated that XR has the ability to manipulate memory and is potentially more powerful than traditional media, they were individual examples of such XRMMs. In this work, our goal is to have a holistic and deeper understanding of all the types of XRMMs that are possible and discuss their potential impact on the user.

\section{Method}
To broadly explore potential XRMMs, we ran three speculative design workshops, bringing together XR and memory researchers. As XR has not yet become a prevalent technology like smartphones, speculative design methods allowed us to explore XRMMs and reflect on their ethical implications, without being restrained to the current usage and capabilities of XR. In this workshop, \participants\ were asked to individually generate XRMM scenarios. These scenarios are grounded in experts' knowledge about XR and human memory, and in the background that was given (section~\ref{sec:bkg}). In the discussion phase, we asked \participants\ to select the most promising scenarios and to discuss the likelihood of such scenarios and their potential outcomes (25 scenarios discussed, \revised{presented in {Table~\ref{fig:scenarios}}}). Finally, we applied thematic analysis to uncover underlying themes and mechanisms used.

\subsection{Participants}

\begin{table}[t]
\resizebox{\columnwidth}{!}{%

\begin{tabular}{lccccc}\toprule
 & ID & Age & Gender & Profession & Years in Profession \\
\midrule

Session 1 & PXR1 & 25 & Female & HCI researcher                  & 1    \\                                               
                           & PXR2 & 28 & Male   & HCI researcher                  & 5                                                   \\ 
                           & PM1  & 24 & Female & Cognitive engineer              & 3                                                   \\ 
                           & PM2  & 24 & Female & Cognitive psychology student    & 4                                                   \\ \hline
Session 2 & PXR3 & 33 & Male   & HCI designer                    & 10    \\                                              
                           & PXR4 & 33 & Male   & HCI researcher                  & 4                                                   \\ 
                           & PM3  & 26 & Female & Cognitive psychology researcher & 3                                                   \\ 
                           & PM4  & 30 & Male   & Cognitive psychology researcher & 8                                                   \\  \hline
Session 3 & PXR5 & 27 & Male   & HCI researcher                  & 6   \\                                                
                           & PXR6 & 27 & Male   & HCI researcher                  & 5                                                   \\ 
                           & PM5  & 38 & Male   & Cognitive psychology researcher & 2                                                   \\ 
                           & PM6  & 30 & Male   & Cognitive psychology researcher & 2  \\   

\bottomrule
\end{tabular}
}
\captionof{table}{Demographic of experts in XR and cognitive psychology}
\Description{This table gives the demographic data of the workshop participants (ID to reference them in the paper, age, gender, profession and years in profession), for each session of workshop.}
\label{fig:participants}
\end{table}

Twelve \participants\ (age: $M = 28.7, SD = 4$ yrs) participated in our workshops (Table~\ref{fig:participants}). As the goal was to explore scenarios of XRMMs, we recruited two types of \participants: HCI researchers specialized in XR (n=6, called XR \participants), and cognitive psychology researchers studying human memory (n=6, called memory \participants). 
We ran the workshop three times, inviting each time two memory \participants\ and two XR \participants.
This configuration enables to have both a discussion between two experts of the same field, and between experts of different fields. We specifically recruited cognitive psychology researchers who have already used XR, to make sure that they have some basic understanding of what is XR, how it can be used and what new possibilities these technologies offer. 
Most of the memory \participants\ rated their XR knowledge to be at least average and mostly above average (5-point Likert scale, M = 3.8, SD = 0.4). Each group of \participants\ rated their knowledge to be at least above average and mostly very high in their expertise field (M = 4.6, SD = 0.5).
The \participants\ were remunerated 30 euros for taking part in the workshop. Sessions were conducted over zoom using Miro as a brainstorming tool. During the last session, PM6 disconnected half-way in the workshop and could not reconnect. The data that was created until the disconnection was included in the analysis.

\subsection{Workshop Procedure}

\begin{figure}[h]
    \centering
    \includegraphics[width=\linewidth]{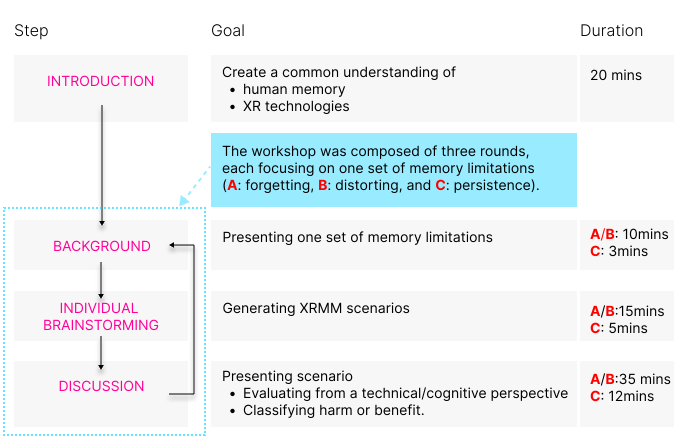}
    \captionof{figure}{ Description of the workshop procedure including steps, goal and duration. The blue square show the three rounds. The red A,B,C point the time for each round.}
    \label{fig:workshop-procedure}
    \Description{This figure shows the procedure of the workshop described in this section. The introduction lasted 20 minutes, and then, there were three rounds (one for each category of memory limitations) of background (10min), individual brainstorming (15min) and discussion (35min).}
\end{figure}

Each workshop session lasted three hours and was composed of four \participants\ and one moderator. It consisted of an introduction, and three rounds of introduction to memory \limitations, individual brainstorming, and discussion (Figure~\ref{fig:workshop-procedure}). 

\subsubsection{Introduction}
To create a common understanding of the different fields, we gave a detailed introduction on human memory (types, process, functions, and \limitations, section~\ref{sec:humanmemory}) and XR technologies (section~\ref{sec:xr}). We warned participants that this background was not exhaustive, and they were invited to refer to their own knowledge and expertise. 

Each round of brainstorming and discussion was focused on one set of memory \limitations\ (forgetting, distorting and persistence, Table~\ref{fig:sevensins}). We chose to separate the workshop sessions into three rounds so that every different type of memory \limitation\ can be equally explored. The order of rounds between workshops was counterbalanced using a latin-square.

\subsubsection{Individual Brainstorming}

\begin{figure}[ht!]
    \centering
    \includegraphics[width=\linewidth]{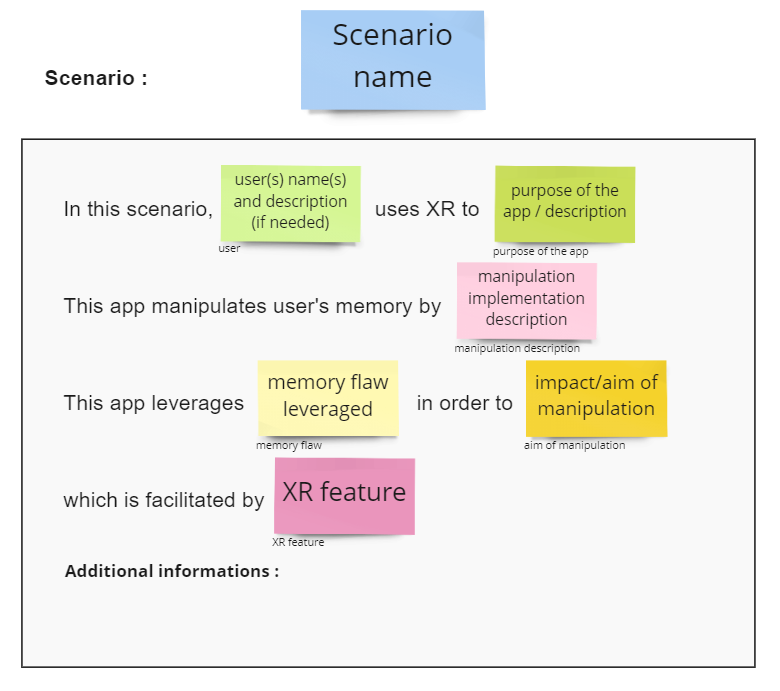}
    \captionof{figure}{Scenario template - Miro board screenshot. This template was used by participants during individual brainstorming sessions to generate scenarios.}
    \label{fig:template}
    \Description{Screenshot of the template that was used by participants during the workshop to create their scenarios. Participants had to fill the parts between brackets: "[Scenario name] : In this scenario, [user(s) name(s)] uses XR to [purpose of the app]. This app manipulates user's memory by [manipulation implementation description]. This app leverages [memory flaws leveraged] in order to [aim of manipulation] which is facilitated by [XR feature].}
\end{figure}

Experts were asked to generate as many XRMM scenarios as possible by filling a template (Figure~\ref{fig:template}). The goal of this template was to help the \participants\ to give more details about the scenarios: who would use the XR application and why? What is the aim of the manipulation and how is it implemented? The \participants\ were free to go beyond that template. 

\subsubsection{Discussion}

During the discussion phases, each \participant\ had two minutes to present a scenario in more depth. \revised{They were instructed to select scenarios for discussion that they deemed to be the most interesting and valuable}. The other \participants\ and moderator could ask questions about the scenario and add comments. Then, each \participant\ rated three aspects of the scenarios (possibility to implement such XRMMs from a (1) technical and (2) cognitive perspective and (3) classify harm or benefit), to trigger discussions.
This phase was recorded to be transcribed. 

\subsection{Data Collection}

\revised{Overall, participants generated 74 scenarios. 25 of them were presented and discussed by the participants during the workshop. We provide a short description of these scenarios in {Table~\ref{fig:scenarios}}. The coding was conducted on the transcripts of these 25 discussions. The remaining 49 scenarios existed as templates filled in the Miro board but could not be explained and discussed in detail during the workshop due to time limitations. As these scenarios were often incomplete and lacked details to be fairly represented, we did not use them for the coding sessions, and we do not provide them in this paper. However, some of these non-discussed scenarios contained enough information to derive high-level constructs from them. These scenarios were used in a final session to count the number of scenarios matching the XRMMs definition ({section~\ref{sec:exclusion})} and to validate the classes found with the coding.}


\subsection{Thematic Analysis}
\revised{To analyze the data, three authors conducted synchronous collaborative coding sessions (7 sessions each taking 4 hours, 28 hours in total). 
In a first meeting, the three coders started by creating an initial codebook (deductive coding) around questions of interest (who is involved in the manipulation, how does it work and why is it used). Afterward, each coder read through all the transcripts of the 25 discussed scenarios without coding them. Then, during 4 meetings, the three coders discussed each scenario. The scenarios were coded following 1) the initial codebook and 2) an open coding procedure (inductive coding). Once a new code that was not covered by the existing codebook came up, the coders discussed this new code, which was added only if all the coders reached an absolute consensus. In the last two sessions, the created codes were structured into themes and connected to larger structures and concepts, resulting in our framework and classes of XRMMs. Conflicts were resolved during the meetings by discussing individual perspectives on every codes and themes until reaching a consensus.}

\subsection{Scenarios Exclusion}
\label{sec:exclusion}
\revised{After the final coding, we derived a definition of XRMMs ({section~\ref{sec:definition}}) which we used to filter scenarios that did not match the definition. This resulted in 3 of the discussed scenarios to be removed (S23, S24, S25, {Table~\ref{fig:scenarios}}), because they do not intend to impact memory, and they are not based on memory limitations. In the scenario S23, advertisements are added in users' field of view to give visibility to brands, but there is no clear intention to distort memories or to influence the cognitive process of memorization. The scenarios S24 and S25 both leverage individual user preferences (i.e., smells and cultural items) to make them appreciate a product, without having an impact on memory.  15 of the non-discussed scenarios did not match the XRMM definition for the same reasons, and 8 did not have enough description to be understood. We obtain a total of 48 scenarios of XRMMs (22 discussed and 26 non-discussed). In the following, the scenarios we refer to are the 22 discussed scenarios of XRMMs (S1 to S22, presented in {Table~\ref{fig:scenarios}}).}

\section{Results}
Based on our initial questions of interest (who is involved in the manipulation, how does it work and why is it used) and the resulting themes and codes, we present a definition of XRMMs, a framework of their core components and three \classes\ of XRMMs.

\subsection{Definition of XRMMs}
\label{sec:definition}
To provide a definition of XRMMs we first needed to provide a general definition of perceptual manipulations, and then show how XRMMs are different from other types of perceptual manipulations, such as VPPMs \cite{tseng2022}.\\     

\textbf{Definition of Perceptual Manipulations:}\emph{ Perceptual manipulations are mechanisms grounded in limitations of users' cognition and perception with a clear intention to influence users towards a specific outcome.}\\

This general definition is classifying PMs as a mechanism (defined by Webster dictionary as ``a process, technique, or system for achieving a result'' \footnote{``Mechanism'' by Webster dictionary:  \url{https://www.merriam-webster.com/dictionary/mechanism}}) and describing the two core components of perceptual manipulations: (a) a grounding in cognitive and perceptual limitations and thresholds and (b) a clear intention to steer towards a specific outcome. The outcome could be effecting change in beliefs, attitudes, intentions and resultant behaviour/responses - or influencing any other cognitive and affective response. We kept this definition open to be able to include all future types of perceptual manipulations. 
Grounded in our analysis of the different \classes\ of XRMMs and their core components, we created the following definition of XRMMs.\\

\textbf{Definition of XR Memory Manipulations:}\emph{ XRMMs are perceptual manipulations grounded in memory limitations with the intention to impact memories, leading to long-term effects on users, through the self, social and directive functions of autobiographic memory.}\\

The intention of impacting users' memory allows us to distinguish XRMMs at the stage of encoding from general perceptual manipulations, and to distinguish XRMMs from VPPMs \cite{tseng2022}.\\

\textbf{Definition of Virtual-Physical Perceptual Manipulations:}\emph{ VPPMs are perceptual manipulations that are grounded in visual-haptic limitations with the intention to nudge the user’s physical movements}\\

\subsection{XRMMs Framework}

\begin{figure*}[t]
    \centering
    \includegraphics[width=14cm]{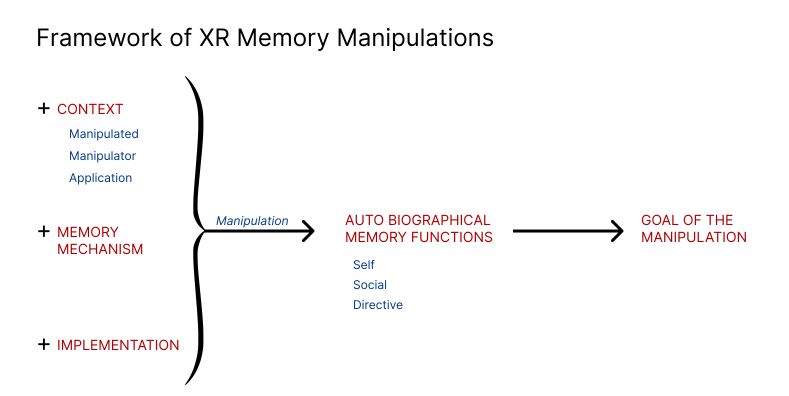}
    \captionof{figure}{Framework presenting the five core components of XRMMs}
    \label{fig:framework}
    \Description{This figure summarizes the five core components of XRMMs presented in this section (context, memory mechanism, implementation, autobiographical memory functions, goal of the manipulation.}
\end{figure*}

We found that five core components were able to describe the intentions and structures of XRMMs:
\emph{the context, the memory mechanisms, the implementation, the memory functions and the final goal} (Figure~\ref{fig:framework}). These components were part of all the generated scenarios. 

\subsubsection{Context}

XRMMs are done in a certain context. There is a \emph{manipulated}, who is the target of the manipulation, a \emph{manipulator}, who is at the origin of the manipulation, and an \emph{XR application} within which the manipulation is occurring.

10 out of 22 scenarios are targeting a vulnerable population (e.g., people suffering from trauma or memory problems). 
This population can benefit from XRMMs (e.g., to treat a trauma), but they can also be harmed.

In 6 out of 22 scenarios, the manipulator and the manipulated were the same person. In this case, the user is manipulating themself through XR, knowing that the results of the manipulation would benefit them. 
When the manipulator is not the manipulated, the scenarios represent three types of manipulators: people in power, businesses, and attackers. These actors are always using XRMMs for their own interest. 

Finally, as XRMMs are based on XR technologies, they are always done within an XR application. In the scenarios that were generated, the XRMMs are done within various types of applications (e.g., training applications, social media, entertainment, therapy tool, etc). The purpose of the application is not necessarily the same as the purpose of the manipulation. In this case, the user is often not aware of the manipulation. 

\subsubsection{Memory Mechanism}
XRMMs are relying on memory mechanisms. In this work we focus on long-term memories, and we grounded the workshops in the seven memory \limitations\ presented by Daniel Schacter (Table~\ref{fig:sevensins}) \cite{schacter2002}. These \limitations\ can either be \emph{leveraged} (taking advantage of memory \limitations\ to manipulate it), or \emph{prevented} (using XR to overcome these \limitations). 14 scenarios are leveraging memory \limitations\ (for both positive and negative purposes), and 8 scenarios are preventing them (for positive purposes only). Each of the seven memory \limitations\ were explored by the scenarios. 3 scenarios focused on transience, 4 on absent-mindedness, 1 on blocking, 7 on misattribution and suggestibility, 3 on bias and 4 on persistence. We found that depending on the moment when the manipulation takes place in relation to the memory process (\emph{at encoding, pre-retrieval, at retrieval}), and depending on the type of memory \limitation\ that is leveraged, different implementations and strategies are used. This constitutes three \emph{\classes\ of XRMMs} which are described in sections \ref{sec:encoding}, \ref{sec:retrieval} and \ref{sec:preretrieval}.

\subsubsection{Implementation}
We found a split between types of implementation (AR and VR techniques) that are used. In AR, XRMMs can be implemented by \textit{adding digital elements to the real world} (e.g., adding a frame around an object), by \textit{removing elements from the real world} (e.g., hiding an object), or by \textit{modifying elements} (e.g., replacing an real object by a different virtual one). In VR, manipulations can be done by \textit{immersing users in a new environment}, or in a \textit{3D reconstruction of a past event} that they already experienced before. This reconstruction can either be accurate or altered. In sections \ref{sec:encoding}, \ref{sec:retrieval} and \ref{sec:preretrieval}, we present what implementations are used depending on the memory mechanisms involved.

\subsubsection{Outcomes}
This context and memory mechanisms, associated with an implementation, can influence the cognitive processes associated with the encoding or retrieval of memory, or influence the perception of past events. This manipulation can \emph{impact users through the three functions of AM} (section~\ref{sec:amfunctions}) \cite{bluck2005}. The scenarios explore how XRMMs can have an influence on self-perception, preferences and emotional state (related to the self-function of AM), on relationships and empathy (related to the social function), and on decisions and behaviors (related to the directive function). 

Impacting users through the memory functions (self, social and directive) enables to achieve a \emph{final goal}. For example, XRMMs could be used to influence decisions (impact through the directive function), with the final goal to make someone buy a product, or it could be used to have an impact on self-confidence (e.g., impact through the self-function), with the final goal to help someone perform better in a task.  From the scenarios, we identified four main types of final goals: purely malicious goals (e.g., creating chaos), making money (e.g., selling a service or a product), improving mental state (e.g., coping with traumas), and self-optimization (e.g., performing better in a task).

\begin{figure*}[ht]
    \centering
    \includegraphics[width=12.5cm]{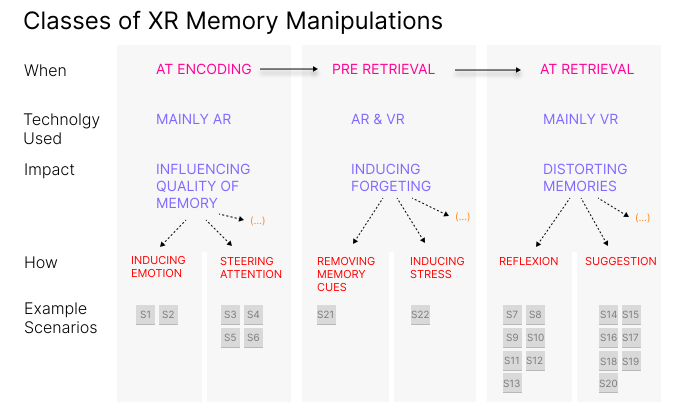}
    \captionof{figure}{The three classes of XRMMs (\emph{at encoding, pre-retrieval, retrieval}), which differ in terms of \emph{technology used},  and \emph{impact} on memory. Two examples of mechanisms (\emph{How, Example scenarios}) are given for each class.}
    \label{fig:classes}
    \Description{This figure summarizes the classes of XRMMs described in this section (when they happen, technology used, impact on memory, how it works). XRMMs at encoding mainly use AR, and influence quality of memory by inducing emotions or steering attention. XRMMs pre-retrieval use both AR and VR, and induce forgetting, by removing memory cues or by inducing stress. XRMMs at retrieval mainly use VR, and distort memories, through reflection or suggestions.}
\end{figure*}

\subsection{Threat Model}

\revised{Grounded in the ``manipulator'' theme of our framework, we derived a threat model outlining future abusive scenarios. An attacker wants to influence a user's future decisions, social interactions, or self-perception ({section~\ref{sec:amfunctions}}), by leveraging memory manipulations in XR (XRMMs). 
To achieve this, the attacker needs access to the software of the device, and to be able to control its function (e.g., access to the user's data, the device's sensor data, and control over device output). 
We outline three routes towards enacting such a threat that emerged from our dataset: }
\begin{description}[wide]
\item[Non-consensual manipulation] \revised{by tricking the VR user into installing malware (e.g., similar to current approaches in smartphones, as done in the Pegasus spyware \footnote{Pegasus, smartphone spyware \url{https://en.wikipedia.org/wiki/Pegasus_(spyware)}}). This can be seen in scenarios S3, S13, S17, S19, and S20, where a hacker gains control over a user's software.}
\item[Exploiting a position of power] \revised{by having access to the device from a position of power and authority (e.g., by being a caretaker of a person using VR as a therapy tool, as in scenarios S7, S8, S10, and S11).}
\item[Consensual, incentivized manipulation] \revised{by gaining the user's permission (e.g., making the user accept to be manipulated in exchange for free services). With the rise of computing technology and social media, we accepted a new form of direct marketing and advertising that is tailored directly toward the individual consumer and leverages psychological principles such as those outlined in dark design patterns. Similar business models could potentially be done with XRMMs (as seen in scenarios S1, S12, S15, and S18). This opens a potential threat around the user willingly accepting a method that is actively influencing their memory as a ``currency'' to get access to a service (similar to today's approach of ``paying'' for social media with our data).}
\end{description}

\subsection{\Classes\ of XRMMs}

After coding each scenario we noticed one set of codes focusing on the moment when the manipulation occurs. Knowing the stages of formation of long-term memories (encoding, storage and retrieval, section~\ref{sec:humanmemory} \cite{melton1963, shiffrin1969}), we were able to identify three moments when a manipulation can occur: while the event is happening (\emph{at encoding}), while it is remembered (\emph{at retrieval}), or between encoding and retrieval (\emph{pre-retrieval}) (Figure~\ref{fig:classes}). All of the generated XRMMs scenarios fall into one of these three categories. In 6 scenarios, the manipulation occurs \emph{at encoding}, 14 others occur \emph{at retrieval}, and 2 others occur \emph{pre-retrieval} \revised{({Table~\ref{fig:scenarios}})}. After analyzing these scenarios, we found three main \classes\ of XRMMs that differ in terms of technology and impacts: 1) manipulations \emph{at encoding} aim at \emph{influencing the quality of memory encoding} and rely mainly on the \emph{capacity of AR to visually alter (add, remove, modify) reality}; 2) manipulations \emph{at retrieval} aim at \emph{distorting the perception of past events} and rely mainly on \emph{immersive reconstruction of scenes in VR}; 3) manipulations \emph{pre-retrieval} aim at \emph{inducing forgetting}, using both AR or VR specific abilities (Figure~\ref{fig:classes}). 

In the following, we present each XRMM class by providing examples of scenarios generated during the workshop and grounding them in cognitive psychology research. For each \class, we present two examples of ways to impact memory (\emph{``How''} in Figure~\ref{fig:classes}). These are just examples of mechanisms that were explored during the workshops, other mechanisms could be explored by future research. To improve readability, we will not only present the \classes\ as results but add an explanation and interpretation as would usually be done in a separate discussion section. Our final discussion will focus on the larger concept of XRMMs and perceptual manipulations.

\section{XRMMs at Encoding}
\label{sec:encoding}

\begin{figure*}[t]
    \centering
    \includegraphics[width=\textwidth]{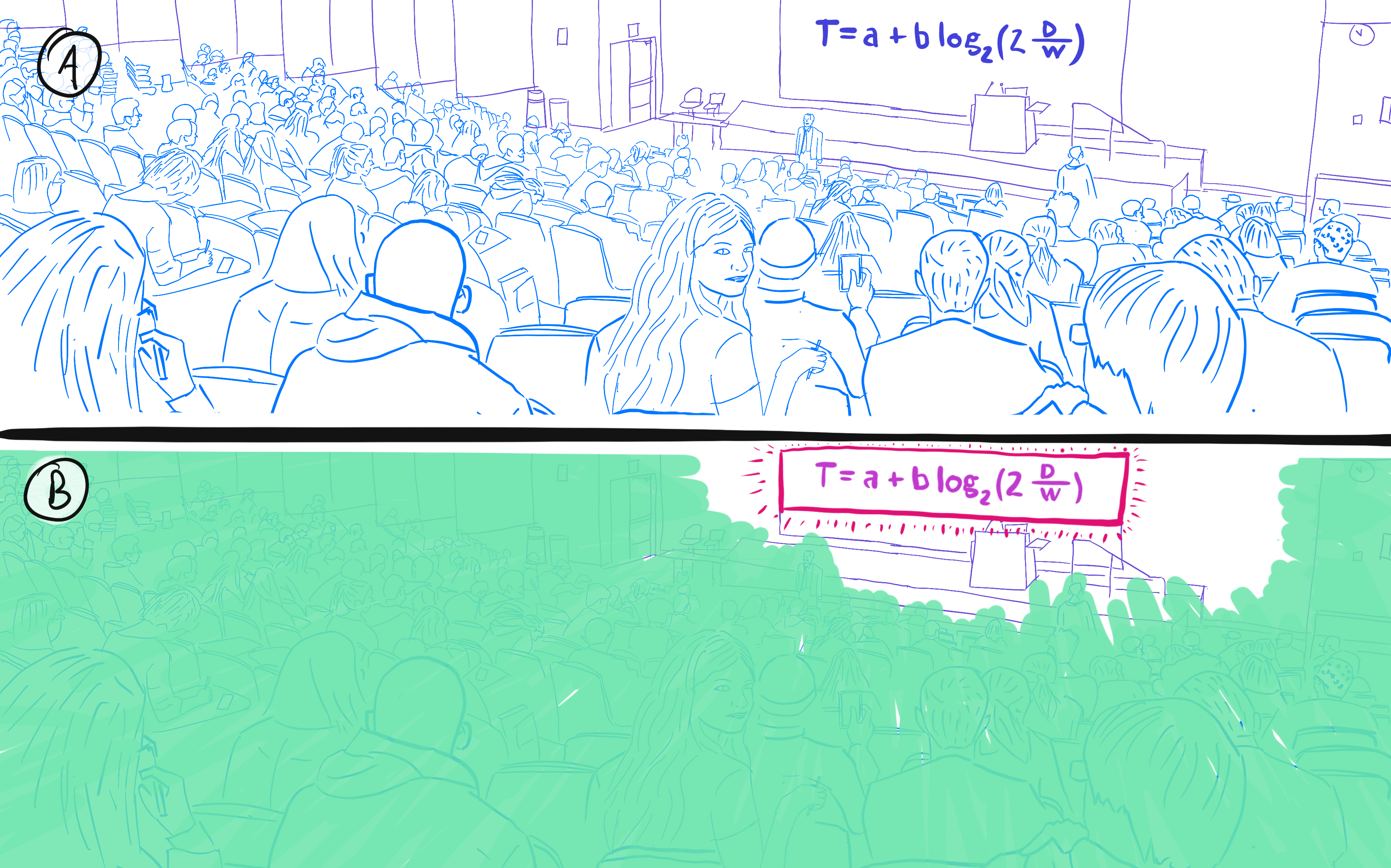}
    \caption[]{Illustration of ``This You Should Remember'' (S6).
	Louise uses AR to better encode important information. Figure
	\textbf{A} illustrates the unaltered field of view, a scene crowded with visual stimuli. Figure \textbf{B} illustrates in  \color{myGreen}{green} \color{black}   how the AR application is occluding the elements that are considered by the system as distractions, and  in \color{myPink}{pink} \color{black}  how information that are considered worth remembering are emphasized.}
    \label{fig:scenario_Remenber}
    \Description{This figure is an illustration of the scenario "This you should remember". In figure A, we see a crowded classroom, with a formula on the teacher's board. In figure B, we see what AR users' would see: this distraction and crowd are hidden, and the formula on the board is highlighted.}
\end{figure*}

The first moment when a memory can be manipulated is \emph{at encoding}, simultaneously as the original perception of the event. The encoding phase is important because it determines what will be stored in long-term memory, and how it will be stored \cite{melton1963}. Therefore, acting at encoding could enable to act on the quality of the memory. From the 6 out of 22 discussed scenarios of XRMMs at encoding, we found that these manipulations could be done \emph{by inducing emotions} or \emph{by steering attention} (by preventing or leveraging absent-mindedness), mainly using AR. We found that XRMMs at encoding could be used to help encode better quality memories, but it could also be used for the opposite purpose: to deliberately make someone's memories weaker. In the following, we explain and discuss these two means to influence encoding: \emph{inducing emotions} and \emph{steering attention}.

\subsection{Influencing Encoding by Inducing Emotions}
Two scenarios (S1, S2, \revised{presented in {Table~\ref{fig:scenarios}}}) explore how \emph{XRMMs could influence encoding by inducing emotions}.
\subsubsection{Grounding}
 Emotions modulate every aspect of cognition, and particularly memory \cite{tyng2017}. It is established that the emotions felt during an event influence the retention and recall of the memory: experiences linked to strong emotion are retained longer, more vividly and accurately \cite{tyng2017, phelps2004, cahill1998}. While some studies indicate that positive emotions facilitate learning \cite{um2012}, negative emotions such as stress or fear can also facilitate or alter the quality of memory encoding, depending on the intensity of the emotion \cite{vogel2016, mcgaugh2002}. 
 
 \subsubsection{Scenarios}
 In the scenario ``You'll Remember the Unlikely Risky Event'' (S2), XR is used to induce strong emotion in a learning context, so the information is better encoded and remembered later (e.g., stimulating an accident while learning how to drive in VR).  Another scenario, ``Peak-End Rule'' (S1), uses XR to induce intense positive emotions, which will be better remembered than the negative ones and improve user's self-perception. 

\subsection{Influencing Encoding by Steering Attention}
Four scenarios (S3, S4, S5, S6, \revised{presented in {Table~\ref{fig:scenarios}}}) explore how \emph{XRMMs could influence encoding by steering attention with AR, preventing or leveraging absent-mindedness}.

\subsubsection{Grounding}
  The encoding of memories is selective: we are constantly exposed to large quantities of stimuli from our senses, but since storing all this information is impossible, we make a first selection at encoding \cite{uncapher2009, cowan1988, desimone1995}. Absent-mindedness (Table~\ref{fig:sevensins}) is when forgetting occurs because insufficient attention is given at encoding (e.g., because of mind wandering or divided attention) \cite{craik1996, baddeley1984, blonde2022, schacter2002}. This can be observed in situations when people forget where they left their glasses, because it is the kind of task that is carried out automatically, without paying enough attention. 

\subsubsection{Scenarios}

``This You Should Remember'' (S6) depicts well how absent-mindedness could be prevented or leveraged. In this scenario, Louise has AR glasses and uses them to offload some cognitive tasks. The AR application chooses for her what is worth remembering, and steers her attention towards it, so it is better encoded. Thanks to the AR glasses’ ability to visually alter reality (adding content, removing or modifying), important elements in the environment are framed or highlighted, and the non-important distractions are blurred (Figure~\ref{fig:scenario_Remenber}). Detecting and deciding which are the important elements is done through an AI process, based on crowd-sourced data. The application is preventing absent-mindedness by helping Louise to focus on important things to be remembered. It is also leveraging absent-mindedness, to limit the encoding of events deemed unworthy. 
Two other similar scenarios were generated (S4 and S5), where inattention is detected using biometric or neurological signals, and attention is directed using visual feedback in AR or by hiding distractions.

\subsection{Discussion}

\subsubsection{\revised{Technological Challenges}}
In scenarios S1 and S2, influencing the quality of memory encoding relies on the capacity of immersive environments to induce emotions. Since VR can induce vivid emotions and reactions similar to real events \cite{felnhofer2015, zimmer2019, martens2019, slater2009}, these manipulations are technically plausible.
The scenarios influencing the quality of memory encoding through attention steering rely on the capacity of AR to visually alter reality. As research focusing on how to alter or diminish reality in real time shows promising results \cite{herling2012,billinghurst2015, cheng2022, mori2017}, such manipulations are plausible in a near future from a technological perspective.
\revised{Another technical challenge for such manipulations is the ability of AI tools to identify in real time what represents a distraction and what is important. However, research in the direction of object detection and identification in images made a lot of progress these last years \mbox{\cite{zhao2019, vinyals2015}}.}

\subsubsection{\revised{Role of XR}}
During the analysis, we additionally coded currently existing techniques that would have a similar impact on the encoding of memory. \revised{We found that influencing memory encoding by inducing emotions is not unique to XR, and scenarios similar to S1 and S2 could be implemented using smartphone applications. However, the capacity of XR to fully immerse the user in an environment allows to induce vivid strong emotions \mbox{\cite{felnhofer2015, zimmer2019, martens2019, slater2009}}, which could amplify this type of manipulation. We also found that the current techniques to steer attention (e.g., noise-canceling headphones or focus mode on smartphones), cannot visually steer users' attention. XR (and AR in particular here) is unique in its ability to visually alter the perception of the real world. Using this could enable to interfere with the encoding process more effectively than most currently known techniques.}

\subsubsection{Purpose}
The generated scenarios suggest that XRMMs at encoding could help and benefit AR users. In four scenarios (S2, S4, S5, S6), the goal of XRMMs is to optimize brain capacities and offload cognitive tasks. PM3 mentioned that such techniques could be used by healthy users but could also help people suffering from Attention Deficit Disorders (ADD). 
However, another scenario (S3) relies on the same idea of influencing encoding through attention, but with the opposite purpose. In this scenario, distractions are added to the field of view of the person wearing AR glasses to make their memory encoding weaker, and make them vulnerable to attacks. While the intentions behind the previous scenarios were positive, this scenario depicts a malicious use of XRMMs at encoding.

\section{XRMMs at Retrieval}

\begin{figure*}[!ht]
    \centering
    \includegraphics[width=\textwidth]{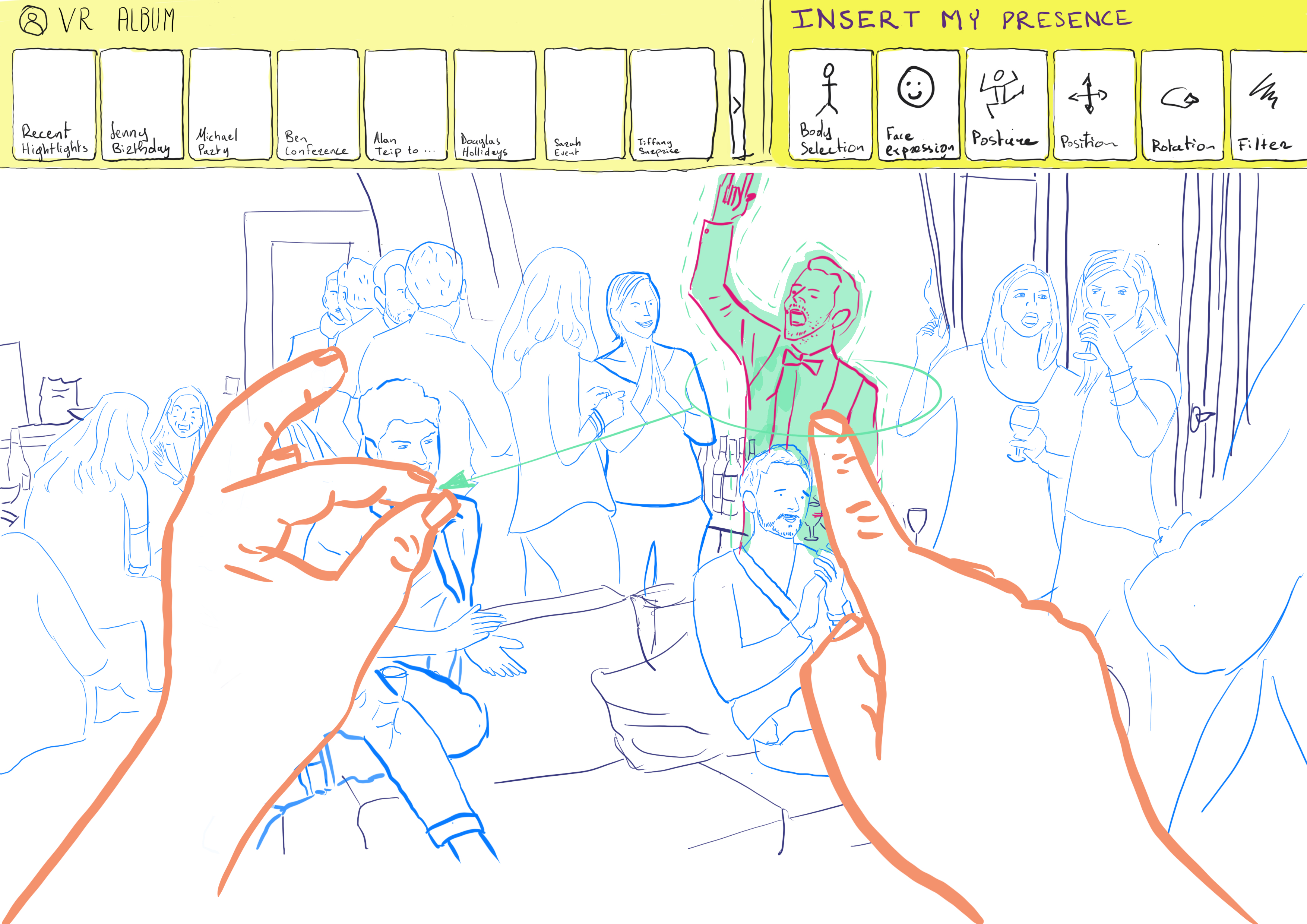}
    \caption[]{Illustration of the scenario ``Buy a Friend'' (S18).
	Jim inserts himself in the VR album of Jenny's party, to distort
	her memory and make her remember he was there. In \color{myBlue}
	{blue} \color{black}  the participants that were present at
	Jenny's party, in \color{myOrange} {orange} \color{black}  the hand of Jim editing the VR reconstruction of the party, in \color{myPink} {pink} \color{black}  the insertion of the 3D capture of Jim inside the party, in \color{myGreen} {green} \color{black}  the altered parts of the reconstruction.}
    \label{fig:scenario_buyFriend}
    \Description{This figure illustrates the scenario "Buy a Friend". The figure is described in caption.}
\end{figure*}

\label{sec:retrieval}
A memory can be manipulated \emph{at retrieval}, while it is being remembered. Memory storage is dynamic: memories are susceptible to change when they are retrieved, and retrieval leads to the formation of new memories \cite{alberini2013, alberini2011, sara2000}. Therefore, acting while a memory is remembered could enable to act on its perception, and distort it. From the 14 discussed scenarios of XRMMs \emph{at retrieval}, we found that these manipulations could be done \emph{using VR reconstructions of past events as a reflection tool} (to leverage positivity bias and prevent persistence), or \emph{using distorted reconstructions} (leveraging misattribution and suggestibility). We found that XRMMs \emph{at retrieval} could be used to influence someone’s perception of the past for various positive purposes (e.g., to prevent suffering), but it could also be used to make someone vulnerable. In the following, we explain and discuss these two means to influence perception of the past and distort memories: \emph{supporting reflection} and \emph{inducing suggestions}.

\subsection{Influencing Perception through Reflection}
Five scenarios (S7, S8, S9, S10, S11, S12, S13, \revised{presented in {Table~\ref{fig:scenarios}}}) explore how \emph{XRMMs could influence perception of past events by using VR as a reflection tool, preventing or leveraging bias and persistence}. 

\subsubsection{Grounding}
Reflection is the process of reviewing past experiences to learn from it and to reframe it. Memories can be actively reviewed through this process \cite{konrad2016}. While reflecting on positive events increases the ability to enjoy life and helps to perceive things more positively \cite{bryant2005}, reflecting on negative events (e.g., being confronted to traumatic events) can help to cope with them on the long-term \cite{pennebaker1986}. Reflection can be done without employing technology by mentally reviewing memories of past personal experiences, or using technologies support, by reviewing rich digital records of past events (e.g., text, images or videos) \cite{konrad2016a, isaacs2013}. 

\subsubsection{Scenarios}
``Therapy'' (S8) depicts how VR could be used to reflect on past experiences and reframe them. In this scenario, a patient is immersed in a VR reconstruction of a past event that they are discussing with the therapist. This reconstruction helps the patient to retrieve the memories of that event, and analyse the situation with the information that they have now but they did not have when the event occured.
``Couple Game'' (S9) and ``Dysmorphia'' (S10) also explore that direction, respectively to help couples communicating, and to correct self-perception. “Traumatic Memory” (S7) shows how a therapist could use VR as a tool to treat a patient suffering from Post-Traumatic Stress Disorder (PTSD), by immersing to them in the context of the trauma. This method, called Virtual Reality Exposure Therapy (VRET), is already explored and shows promising results \cite{sherrill2019, difede2007, gerardi2008}.

However, even if being exposed to traumatic memories can be benefiting on long-term, it can also be harmful on short term.
Two scenarios (S12, S13), explore how malicious actors could expose VR users to traumatic events to make them suffer out of malicious intend or make them uncomfortable with the intention to sell an option which could prevent future exposure.  

\subsection{Distorting Memories through Suggestions}
Seven scenarios (S14, S15, S16, S17, S18, S19, S20, \revised{presented in {Table~\ref{fig:scenarios}}}) explore how \emph{XRMMs could distort memories by using altered 3D reconstructions of past events, leveraging misattribution and suggestibility}.

\subsubsection{Grounding}
Misattribution is when a memory is present in mind but attributed to the wrong source (Table~\ref{fig:sevensins}) \cite{schacter2002, schacter2001}. This effect occurs because a memory is not labeled with its source: it is unconsciously deduced during retrieval, from different cues. If these cues are not precise enough, the right source cannot be retrieved \cite{johnson1981, johnson1993}.  A good example of misattribution is when someone think they remember an event they experienced, when they actually just remember the picture of that event. Suggestibility refers to the fact that memories can be distorted if a new information is added between encoding and retrieval, through leading questions or comments \cite{schacter2002, loftus1995, loftus2005}. It is related to misattribution because distortion from suggestion always involves some misattribution between the real event and the suggested information. 

\subsubsection{Scenarios}


“Buy a Friend” (S18) depicts well how misattribution and suggestibility could be leveraged. In this scenario, Jim uses a VR social media application where immersive reconstructions of past events can be revisited and shared with other users. With his premium account, Jim has the possibility to insert himself in other peoples' memories. As he wants to get closer to Jenny but could not go to her last party, he decides, through the application, to insert a realistic 3D representation of himself in the memories that Jenny shared (Figure~\ref{fig:scenario_buyFriend}). Later, when Jenny revisits her memory in the application, she sees that Jim was there. Because of misattribution (between the reconstructed event and the real event) and suggestibility (integrating the suggested information that Jim was there), Jenny’s original memory might be distorted: “Jim was there too, we had such a nice time together.”. In this scenario, the social media business sells a service that uses XRMMs at retrieval. Six other scenarios (S14, S15, S16, S17, S19, S20) also rely on the similar idea that altered 3D reconstructions of past events could be used to overwrite and distort the memory of the original event.

\subsection{Discussion}

\subsubsection{\revised{Technological Challenges}}
\revised{The main technological challenge of the scenarios of XRMMs \emph{at retrieval} is the ability to make 3D captures of events.} As volumetric capture techniques are already being developed and show good results \cite{lindlbauer2018, fender2022}, this is plausible from a technological perspective. \revised{The manipulations aiming at distorting memories imply the ability to alter the 3D reconstructions of past events (e.g., inserting someone in a VR memory, S18). This could be done manually with some time and efforts, but represents a technological challenge to be automated. However, recent advances with inpainting methods \mbox{\cite{newson2013a, qin2021}} and generative AI tools, such as Stable Diffusion \mbox{\cite{zotero2142}} or DALL.E 2 \mbox{\cite{zotero2140}}, suggest that this would be plausible in a near future.}

\subsubsection{\revised{Role of XR}}

\revised{Influencing the perception of the past through reflection is not unique to XR, as it can be done without technological support, or by using pictures and videos \mbox{\cite{konrad2016a, isaacs2013}}. However, VR represents a more attractive way to reminisce \mbox{\cite{xu2020}} and enhances episodic memory performance \mbox{\cite{makowski2017, smith2019}}, which could facilitate this type of manipulation. Distorting memories through suggestions} could already be done using pictures (e.g., the scenario “Buy a Friend” (S18) could be done by inserting people in pictures shared on social media) \cite{risko2019,nash2009, wade2002}, but the realism and embodiment enabled by XR could amplify the misattribution effect. In the Reality Monitoring framework, Johnson and Raye identify the cues used at retrieval to differentiate between external (real) and internal (imagined) events \cite{johnson1981, johnson1993}. Memories that have a lot of sensory characteristics (e.g., tactile, olfactory), semantic content (e.g., detailed colors, shapes) and contextual information (e.g., spatial, temporal) are more likely to be memories of external events. However, an internal event that has plenty of characteristics corresponding to an external event (e.g., rich in perceptual details) is likely to be confused with an external event. The Reality Monitoring framework can be combined with media richness theory to explain why one could assume that increasing the ``richness'' (e.g., modalities, visual quality) of VR experiences might also lead to being more vulnerable to misattributions between real and virtual experiences \cite{segovia2009, hoffman2001, hoffman1995, fernandes2015, rubo2021}. Segovia et al., were able to demonstrate that using VR to induce false memories in elementary school children is more effective compared to the baseline condition \cite{segovia2009}. \revised{Although distorting memories can be done with more traditional media like pictures, using VR to induce suggestions could be more efficient and amplify the effect of the manipulation.}

\subsubsection{Purpose}
This \class\ of manipulation can serve various purposes. Immersive reconstructions of past events could be used as a reflection tool to understand the past better and reframe it more positively to prevent suffering (e.g., S8).  PM1 suggests that such manipulations should only be done by a professional third party (e.g., a therapist), as immersing users in reconstruction of traumatic memories could also be harmful (e.g., S13, S12).

Even though in two scenarios (S14, S16) altering reconstructions of past events benefits the user (e.g., in S14, a VR user distorts their own memories to have a better impression about their past and increase their confidence), in three other scenarios (S15, S17, S20), the XRMMs serve malicious purposes (e.g., selling something from deception or making someone feel incompetent). In the scenario ``Buy a Friend'' (S18), the initial intention is positive, as the goal of the manipulation is to help someone make friends. However, this friendship would be based on false memory induction, which is not ``ethically great'', accoding to PM3. Furthermore, this manipulation could be used to get close to someone with malicious intentions (e.g., to get money from them). These scenarios raise ethical concerns about the possibility to alter the 3D reconstruction of past events.

\section{XRMMs Pre-Retrieval}

\begin{figure*}[ht!]
    \centering
    \includegraphics[width=16cm]{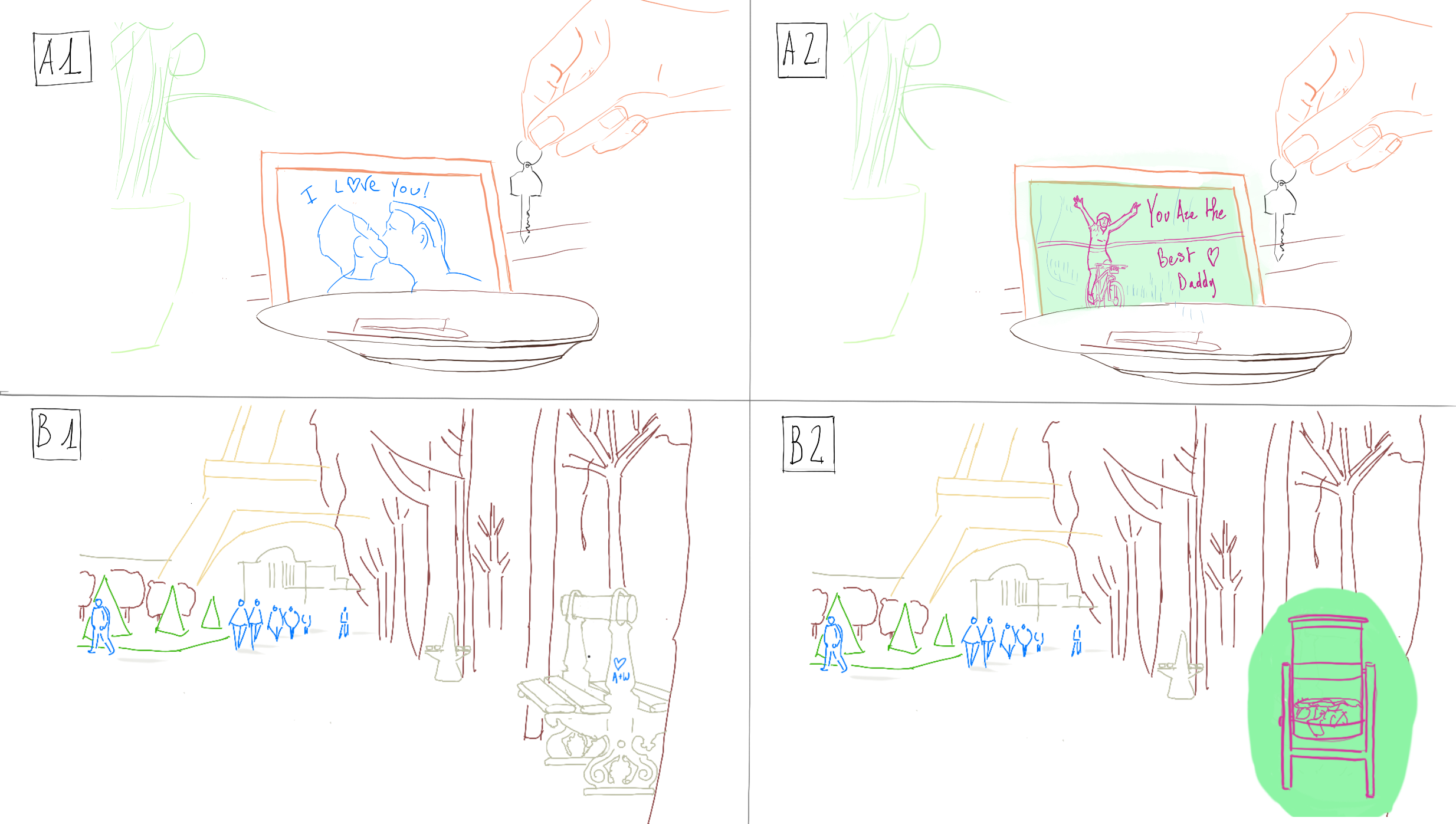}
    \caption[]{Two illustrations of the scenario ``\textit{AR
	Eraser}'' (S21). Alex uses AR to hide the elements associated
	with their former partner, to forget them faster. Figures \textbf{A1} and \textbf{B1} show the unaltered situation. In \textbf{A1} a photo of Alex and his former partner, and in \textbf{B1}, the bench where they kissed for the first time. \textbf{A2} and \textbf{B2} shows what Alex sees with the AR glasses, replacing the elements that might reinforce the unwanted memory. In \color{myPink} {pink} \color{black}  are the replaced elements, in \color{myGreen} {green} \color{black}  are the visual area impacted by the alteration. }
    \label{fig:scenario_arEraser}
    \Description{This figure illustrates the scenario "AR Erase". The figure is described in caption.}
\end{figure*}

\label{sec:preretrieval}
Finally, a memory can be manipulated \emph{pre-retrieval}: after the event occurred, but before it is remembered. From the two discussed scenarios of XRMMs \emph{pre-retrieval}, we found that these manipulations could \emph{induce forgetting}, either by \emph{hiding memory cues} (leveraging transience), or by \emph{inducing stress} (leveraging blocking). Inducing forgetting could be done for positive or negative purposes (e.g., to help someone forget a negative experience, or to make them feel incompetent). As XRMMs pre-retrieval are described in only two of the discussed scenarios, this direction should be explored in more depth. In the following, we explain and discuss these two means to induce forgetting: \emph{hiding memory cues} and \emph{inducing stress}.

\subsection{Inducing Forgetting by Hiding Memory cues}
One scenario (S21, \revised{presented in {Table~\ref{fig:scenarios}}}) explores how \emph{XRMMs could induce forgetting by hiding memory cues, leveraging transience}. 

\subsubsection{Grounding}
Transience refers to a weakening or loss of a memory over time (Table~\ref{fig:sevensins}) \cite{schacter2002}. Memories from a few minutes ago are naturally more precise and vivid than memories from years ago \cite{ebbinghaus18851885}. Transience can be reduced by retrieving memories often (e.g., thinking or talking about it). Memories that are rarely or never recalled will be forgotten more quickly \cite{schacter2002,thompson2013,bahrick2000}. 

\subsubsection{Scenario}

The scenario “AR Eraser” (S21) explores how AR could be used to leverage transience and help someone to forget an unwanted memory. In this scenario, Alex uses AR glasses to hide in their environment all the elements associated with their former partner, which they want to forget faster. These elements could either be blurred (e.g., blurring the bench where the couple had their first kiss), or replaced (e.g., replacing a picture by another one) (Figure~\ref{fig:scenario_arEraser}). By hiding memory cues linked to that person, the thoughts linked to them could be less frequent and the memories could degrade faster.

\subsection{Inducing Forgetting by Inducing Stress}
One scenario (S22, \revised{presented in {Table~\ref{fig:scenarios}}}) explores how \emph{XRMMs could induce forgetting by inducing stress, leveraging blocking}.

\subsubsection{Grounding}
Blocking is when a memory is encoded and stored in mind but cannot be retrieved when needed (Table~\ref{fig:sevensins}) \cite{schacter2002}. An example of blocking is the “tip of the tongue” state, when we cannot remember a word, but we can produce partial information about it, like its first letter. These retrieval failures can occur due to stress \cite{wolf2017,klier}. 

\subsubsection{Scenario}
The scenario “False Training App” (S22) explores how blocking could be leveraged to make VR users vulnerable. In this scenario, Paul uses a VR application that proposes to test his knowledge (e.g. a quiz application). This application immerses users in a stressful environment (e.g., showing a crowd around the user) to intentionally trigger blocking and make them fail the test and feel incompetent. The application then proposes the user to buy a training application, to gain more competence. 

\subsection{Discussion}
\subsubsection{\revised{Technological Challenges}}
The ``AR Eraser'' (S21) scenario relies on the ability of AR to diminish reality and hide elements, which is already explored and partially plausible from a technological perspective \cite{herling2012,cheng2022, mori2017}. \revised{This manipulation also requires AI-backed tools to identify and select the elements which need to be altered (in this scenario, the elements related to a former partner), and to alter them seamlessly. This represents technological challenges, but research about object detection in images \mbox{\cite{zhao2019}}, inpainting \mbox{\cite{qin2021}}, and generative AI \mbox{\cite{zotero2140, zotero2142}} shows promising results.}

\subsubsection{\revised{Role of XR}}
\revised{Although the technique used in the scenario ``AR Eraser'' (S21) to induce forgetting} is already physically feasible without AR for personal objects (e.g., storing every object related to a person in a box out of the user's view, \revised{or removing someone from a picture), AR is unique in its ability to} hide elements that the user cannot physically remove (e.g., a bench on the street), \revised{which makes this manipulation more complete and effective.}

The ``False Training App'' (S22) scenario relies on the capacity to induce stress. It could already be implemented on smartphone applications (e.g., by adding stress factors usually used in games), but VR allows to fully immerse the user in a stressful environment, which can trigger realistic physiological reactions \cite{zimmer2019, martens2019, kotlyar2008}. \revised{This specificity of VR could amplify the blocking effect and make this manipulation easier to induce.}

\subsubsection{Purpose}
The intention behind the scenario ``AR Eraser'' (S21) is positive, as the goal is to prevent suffering caused by painful memories. However, this technique is questionable, as interfering with natural coping processes could have undesirable effects. PXR2 points out that bad memories ``also exist for a reason'', so avoiding them is not necessarily the right solution. 

In the scenario ``False Training App'' (S22), forgetting is induced for malicious purposes (i.e, selling something from deception). This raises ethical concerns about the ability of immersive technologies to induce realistic physiological reactions such as stress.

\section{General Discussion}

Our main goal with this research was to explore what types of memory manipulations that could become possible in XR (\classes), and how the technology is functioning as an enabler or amplifier in these manipulations (framework of core components).

From the thematic analysis of the generated scenarios, we found \textbf{that XRMMs can be described by five main components}: the context, the memory mechanisms, the implementation, the memory functions and the final goal. While the context, implementation and final goal are elements found in most other XR applications, the grounding in memory mechanisms and the impact on users through memory functions is a unique property of XRMMs. Similar to perceptual manipulations, the designer of the application needs to understand the limits of human cognition to implement the manipulation and reach the desired goal. However, with memory manipulations, the designer must not only know how to leverage memory limitations, but must also know how they can impact users through the AM functions (self, social and directive functions) to achieve the desired outcome. This indicates that these XRMMs will require a deeper understanding of human cognition and perception during the design and development process. 

Additionally, we found \textbf{three \classes\ of XRMMs} that are substantially different from each other in terms of technology used (AR vs. VR) and impact on memory (influencing quality of memories, distorting memories, and inducing forgetting). XRMMs \emph{at encoding} rely dominantly on AR technology to interfere with memory building. This \class\ of XRMMs is strongly grounded in the abilities of XR to sense the user and the environment and display digital information over the physical world. XRMMs \emph{at retrieval} rely dominantly on VR technology and the ability to immerse users in 3D reconstructions of past events, to either change their perception of a memory or to distort it. While these manipulations could already be implemented with 2D media such as images, VR is a richer media (relating to media-richness theory \cite{segovia2009, hoffman2001}), and the encoding processes of VR events are closer to those of real events \cite{smith2019, kisker2021}, which could amplify the effect of these manipulations. XRMMs \emph{pre-retrieval} focus on inducing forgetting by hiding memory cues (in AR) or by inducing stress (in VR). This class falls into the same arguments about being potentially more severe than 2D equivalents and enabling new opportunities for impacting users’ memory. 

Working with experts from both fields (XR and cognitive science) allowed us to not only uncover these \classes\ and mechanisms but to ground our speculative design method in existing research about human memory. During the discussion of each workshop, \participants\ were able to comment on the likelihood of scenarios but also discuss what potential positive and negative impacts such scenarios might have on the individual and society. In the following, we present a discussion around potential positive and negative directions which are partially coming out of the discussion of the workshop and partially from our own reflection about our work and prior literature. 

\subsection{The Good: One Step Closer to Augmenting the Human Intellect}

13 out of the 22 scenarios created and discussed in our workshop were focused on positive applications, either to support a neurodivergent population or to augment the abilities of neurotypical users. Both of these could be seen as the augmentation of the human and in particular the human intellect \cite{engelbart1962augmenting}. Douglas Engelbart presented a conceptual framework focusing on increasing a user’s capability to solve complex problems. He gives a set of examples of what these capabilities might be: \emph{more-rapid comprehension, better comprehension, the possibility of gaining a useful degree of comprehension in a situation that previously was too complex, speedier solutions, better solutions, and the possibility of finding solutions to problems that before seemed insoluble}. These capabilities are partially referring to the directive function of memory. By using XRMMs \emph{at encoding}, a user would have \emph{the possibility of gaining a useful degree of comprehension in a situation that previously was too complex}. Steering the user’s attention (e.g., ``This You Should Remember'' S6) or inducing emotional stimuli (e.g., ``You’ll Remember the Unlikely Risky Event'' S2) strengthens the memory of the experienced event and can later lead to better performances due to the directive function of memories. The particularly interesting part of this type of augmentation, compared to Engelbart’s vision, is that XRMMs are not an external tool that is used to perform a task better. XRMMs are a tool to alter perception and memory and thereby change the self, for better or for worse. The main difference is that once the tool is removed, the augmentation still persists, unlike the examples given in ``Augmenting the Human Intellect''.
This is also a core difference between perceptual manipulations such as VPPMs and XRMMs. XR is presenting a vehicle for these types of manipulations since it is now possible to apply them to the real world and not only on digital information. However, this interference with the human perceptual and cognitive system comes with the price of the loss of \emph{Perceptual Integrity}.

\subsection{The Bad: One Step Closer to Losing Perceptual Integrity}

The concept of \emph{Perceptual Integrity} came up during the joint coding session and the analysis of the free discussions that \participants\ had during the workshops. When discussing the scenario S6 where an AI selects what the user should be attentive to (Figure~\ref{fig:scenario_Remenber}), PXR5 mentioned that if ``what is worth remembering is crowdsourced'', users would ``lose some part of their individuality'', which might be ``harmful for democracy and for thinking on our own''. Realizing this distinguishment between a temporal augmentation through a tool and a permanent augmentation through something like XRMMs, we discussed the idea that XR is capable of (sometimes even unnoticeably) interfering with our perception of the real and virtual world and thereby undermining our agency and trust in what we are perceiving to be unaltered by technology (perceptual integrity).

This is particularly possible once the human cognitive system is becoming part of the application design and once the developer can leverage known thresholds to steer users towards actions or beliefs through alterations of their subjective reality. The specific danger we see with XRMMs compared to perceptual manipulations such as VPPMs (e.g., redirected walking \cite{sun2018}) is that XRMMs have a longer-term impact on users. Where redirected walking is leveraging a threshold of the visual system to have an immediate impact on users’ physical movements, XRMMs are leveraging weak spots in our cognitive system that are able to impact the users' self-perception, their social interactions, and decisions and behavior, through the three AM functions (self, social and directive, section~\ref{sec:amfunctions}).  

These findings are building on the works of prior researchers that started to have similar concerns around the ability of XR to influence and manipulate users perception. Roesner et al., presented in 2014 \cite{roesner2014SecuritySystems} and 2021 \cite{roesner2021security} a set of potential challenges around security and privacy of AR technology. They refer to the construct of perceptual manipulations as ``output security'' and started to propose structural changes from a security research's perspective on how to enable a safe \cite{lebeck2016safely} and secure \cite{lebeck2017securing} augmentation. Similar to our interdisciplinary research approach, Baldassi et al., combined neuroscience and computer science in their exploration of the ``output challenge'' and presented one of the core challenges to the ability of AR to manipulate the users sensory, perceptual and cognitive system \cite{baldassi2018challenges}. Our findings are adding to this research directions by presenting a more detailed definition of perceptual manipulations in general and a distinguishment between VPPMs and XRMMs. 

Distinguishing these two types of perceptual manipulations in XR (VPPMs and XRMMs) allowed us to discover and articulate the concept of perceptual integrity. XR as a technology mediates our perception of reality. Consequently it strongly influences our perceptual integrity, i.e., the extent to which our perception of reality is sound and accurate. Approaches that manipulate or undermine perceptual integrity using XR have thus far done so with the intent to influence free will, agency and autonomy - impacting our behaviour, actions and reactions in the process. XRMMs and VPPMs are two powerful examples of the potential for manipulating perceptual integrity using XR, and exemplify the new risks posed to perceptual integrity, particularly by the advent of everyday consumer AR. We argue that both types of perceptual manipulations are only working since they are breaking with the idea of perceptual integrity. However, one could argue that VR and AR are at their core manipulating the perception of the user (e.g., visually displaying digital content that does not exist in the physical world). 

In addressing perceptual manipulations and their impact on perceptual integrity, our research prompts a breadth of future research questions that researchers need to urgently address before we arrive at the point of ubiquitous XR. In particular: How should we balance perceptual integrity against exploiting the advantages and capabilities that XR-driven perceptual mediation unlocks? To what degree do we allow perceptual manipulations to be applied? What mechanisms are required to make users aware of potential or on-going perceptual manipulations without breaking with the experience? What new types of perceptual manipulations will XR enable, and what opportunities, risks, and concerns do they pose? And perhaps most fundamentally of all, who or what should be able to access and utilize this capacity for perceptual manipulation, and under what circumstances is this regulated by law? Such capabilities can empower individuals - but they could also be used by other entities, from businesses to governments, eager to influence or steer the beliefs, attitudes, intentions and actions of users. 

These are all questions that are currently similarly asked in the field of Neurotechnology (technology that interfaces with the users neural system and is able to monitor or modulate neural activities). The Neurorights Foundation presented a set of five rules (Mental Privacy, Personal Identity, Free Will, Fair Access to Mental Augmentation and Protection from Bias) \cite{zotero2097}. While all of them do similarly apply to perceptual manipulations, the rule of Free Will (``Individuals should have ultimate control over their own decision making, without unknown manipulation from external neurotechnologies.'') is the best demonstration that perceptual manipulations in XR have similar concerns and should work on a similar set of rules that try to prevent abuse without constraining the powerful abilities of XR. Slater et al., raises similar concerns around the power that realism has in virtual environments \cite{slater2020ethics}.

\subsection{The Ugly: Who is Responsible for the Trajectory and what can the HCI Community do?}

An important question that we want to address is about the responsibility and trajectory of perceptual manipulation research. In recent years, perceptual manipulation techniques have been frequently published in the field of HCI and XR, mainly uncovering interaction techniques that could leverage users’ perceptual thresholds \cite{razzaque2002,steinicke2010,sun2018,azmandian2016}. While researchers are starting to explore potential harms that could arise from XR and perceptual manipulations \cite{tseng2022,mhaidli2021,ethicalhmd}, these works are mainly arising from the Safety, Security and Privacy community \cite{baldassi2018challenges,lebeck2016safely,lebeck2017securing,roesner2021security,roesner2014SecuritySystems}. XR and HCI researchers are still dominantly focusing on the potential benefits of perceptual manipulations and working towards understanding and quantifying new thresholds. A good example is the field of locomotion and redirected walking in Virtual Reality \cite{razzaque2002,steinicke2010,sun2018}. The majority of publications are focusing on even more precisely detecting the perceptual thresholds of visual rotations. This leads us to the second part of this question: Who is responsible and what can HCI research contribute? 

With this paper, we wanted to demonstrate how we imagine future research around perceptual manipulations. We presented a set of weak spots in human cognition that have the potential to be leveraged in XR for positive and negative application scenarios. The emphasis is on having a holistic perspective of potential threats arising from this type of application. The abusive scenarios were not an afterthought but an essential part of this research. Instead of focusing on just the positive applications XRMMs could enable (augmenting the human intellect), we present a more general exploration, including and reflecting on the potential abusive scenarios (e.g., perceptual integrity). 
With this paper, we want to present how these four important fields (HCI, XR, Security, and Cognitive Science) need to come together and present a critical reflection of their contributions to the scientific field of XR perceptual manipulations.

\section{Limitations}
Using speculative design workshops comes with the risk of generating scenarios that are more fictional than grounded in our current understanding of a subject. To be able to explore such a construct as XRMMs, we needed to apply this method since the distribution XR consumer devices is not yet at the point where this becomes possible. However, to still create a strong foundation in the two respected fields (XR and Cognitive Science) we recruited experts from the individual fields and build our workshop around known and existing memory \limitations. Additionally, we presented each \class\ of XRMM by giving examples of memory mechanisms which could be leveraged, grounded in cognitive psychology research. These examples are not exhaustive and could be extended by future research about XRMMs. 
Since we used a collaborative coding approach, we were not able to calculate Inter-Coder Reliability (ICR) measures during the thematic analysis process. Independent coding and ICR would have provided a more objective measure of agreement. However, our collaborative approach allowed to reach agreement through discussion and initiated an early reflection on the data and the codebook. 
\revised{Also, most of the experts participating in the workshop were participating from a Western European country (11 out of 12 participants), but we did not collect any formal information about their cultural background. On this basis, we cannot claim any cultural diversity in the scenarios generated.}

\section{Conclusion}

In this work we explored how XR could be used to manipulate memory. We ran three speculative design workshops (n=12), with XR and memory researchers creating 48 XRMM scenarios. Through thematic analysis we arrived at a definition of XRMMs, distinguishing them from other types of perceptual manipulations (e.g., VPPMs). Additionally, we presented a framework of their core components, and discussed three \classes\ of XRMMs (\emph{at encoding, pre-retrieval, at retrieval}). To demonstrate the plausibility of our classes we presented a grounding of each in current XR and memory research. The results of this paper allowed us to classify a new type of perceptual manipulation that has long-term impact on users. We present a discussion about the benefits and risks of this type of perceptual manipulation, taking us a step closer to augmenting the human intellect - but at the cost of jeopardizing perceptual integrity.


\begin{acks}
This work was partly supported by French government funding managed by the National Research Agency under the Investments for the Future program (PIA) grant ANR-21-ESRE-0030 (CONTINUUM).
We thank all the workshop participants for their contribution, as well as the anonymous reviewers for their valuable feedbacks.
\end{acks}

\vfill\null
\bibliographystyle{ACM-Reference-Format}
\bibliography{references}

\newpage
\appendix

\begin{table*}[ht]
\centering
\centering
\resizebox{\textwidth}{!}{

\begin{tabular}{ccccc}
\hline
Id                                & Name                                                                          & Author         & Description                                                                                                                                                                                                                                                                       & Class                                                                                                                \\ \hline
S1                                & Peak-End Rule                                                                 & PXR2           & \multicolumn{1}{c|}{\begin{tabular}[c]{@{}c@{}}Positive emotions are induced during the XR experience, to increase positive memories and overall \\ happiness. \revised{This manipulation is implemented by the developers to make users come back to the app.}\end{tabular}}     & \multirow{2}{*}{\begin{tabular}[c]{@{}c@{}}At encoding\\ Influencing encoding \\ by inducing emotions\end{tabular}}  \\ \cline{1-4}
S2                                & \begin{tabular}[c]{@{}c@{}}You'll Remember \\ the Unlikely Event\end{tabular} & PM1            & \multicolumn{1}{c|}{\begin{tabular}[c]{@{}c@{}}Risky and stressful events are simulated when learning how to drive, so the important \\ lessons are better remembered. \revised{This manipulation is a feature of the training app.}\end{tabular}}                                &                                                                                                                      \\ \hline
S3                                & Make Forget                                                                   & PXR4           & \multicolumn{1}{c|}{\begin{tabular}[c]{@{}c@{}}\revised{An attacker add distractions in other user's AR filed of view,} to make them less attentive and forget \\ what they just did, to make them vulnerable to other attacks \revised{(e.g., stealing money)}.\end{tabular}}    & \multirow{4}{*}{\begin{tabular}[c]{@{}c@{}}At encoding\\ Influencing encoding \\ by steering attention\end{tabular}} \\ \cline{1-4}
S4                                & Attention Detector                                                            & PM3            & \multicolumn{1}{c|}{\begin{tabular}[c]{@{}c@{}}\revised{An AR app helps users with Attention Deficit Disorder (ADD) to detect and notify them when}\\ \revised{they are not attentive, to increase the chance of correctly encoding events.}\end{tabular}}                        &                                                                                                                      \\ \cline{1-4}
S5                                & Encoding Enhancer                                                             & PM4            & \multicolumn{1}{c|}{\begin{tabular}[c]{@{}c@{}}\revised{An AR app helps users in a learning context by blurring distractions and contextual information}\\ \revised{(sounds, objects, movements...), to help them focus and enhance encoding.}\end{tabular}}                      &                                                                                                                      \\ \cline{1-4}
S6                                & \begin{tabular}[c]{@{}c@{}}This You Should \\ Remember\end{tabular}           & PXR5           & \multicolumn{1}{c|}{\begin{tabular}[c]{@{}c@{}}\revised{An AR app proposes to support encoding, through an} AI which selects what users should remember, by \\ highlighting important elements and blurring distractions (fig.\ref{fig:scenario_Remenber}).\end{tabular}}         &                                                                                                                      \\ \hline
S7                                & Traumatic Memory                                                              & PM2            & \multicolumn{1}{c|}{During a therapy, patients suffering from PTSD are exposed to their trauma in VR, to reduce flashbacks.}                                                                                                                                                      & \multirow{7}{*}{\begin{tabular}[c]{@{}c@{}}At retrieval\\ Influencing perception \\ through reflection\end{tabular}} \\ \cline{1-4}
S8                                & Therapy                                                                       & PM2            & \multicolumn{1}{c|}{\begin{tabular}[c]{@{}c@{}}During a therapy, patients are immersed in the context of a past event in VR, to help them analyse\\  the situation and understand the past better.\end{tabular}}                                                                  &                                                                                                                      \\ \cline{1-4}
S9                                & Couple Game                                                                   & PXR3           & \multicolumn{1}{c|}{\begin{tabular}[c]{@{}c@{}}\revised{A VR game allows couples to experience each other memories, }\\ \revised{to help them understand each other better.}\end{tabular}}                                                                                        &                                                                                                                      \\ \cline{1-4}
S10                               & Dysmorphia                                                                    & PM3            & \multicolumn{1}{c|}{\begin{tabular}[c]{@{}c@{}}During a therapy, patients suffering from dysmorphia can revisit their memories in VR\\ to correct their self-perception thanks to embodiment.\end{tabular}}                                                                       &                                                                                                                      \\ \cline{1-4}
S11                               & Memory Training                                                               & PM4            & \multicolumn{1}{c|}{\begin{tabular}[c]{@{}c@{}}\revised{A VR app proposes to insert users' relatives in VR entertaining content}\\ \revised{to help users suffering from dementia to remember their relatives.}\end{tabular}}                                                     &                                                                                                                      \\ \cline{1-4}
S12                               & Shame Reminder                                                                & PXR4           & \multicolumn{1}{c|}{\begin{tabular}[c]{@{}c@{}}An AR app reminds users' of memories linked to places. \revised{This app makes profit by reviving }\\ \revised{traumatic memories, because users have to pay to stop being reminded of them.}\end{tabular}}                        &                                                                                                                      \\ \cline{1-4}
S13                               & Naked Man VR                                                                  & PXR5           & \multicolumn{1}{c|}{\begin{tabular}[c]{@{}c@{}}\revised{An attacker inadvertently exposes VR users to reconstruction of traumatic events,}\\ \revised{to harm their mental state and make them vulnerable.}\end{tabular}}                                                         &                                                                                                                      \\ \hline
S14                               & Speech                                                                        & PM1            & \multicolumn{1}{c|}{\begin{tabular}[c]{@{}c@{}}\revised{A VR speech training app proposes to re-experience previous speech performances, through 3D }\\ \revised{reconstructions. These reconstructions are positively altered, to enhance user's self-perception.}\end{tabular}} & \multirow{7}{*}{\begin{tabular}[c]{@{}c@{}}At retrieval\\ Distorting  memories\\ through suggestions\end{tabular}}   \\ \cline{1-4}
S15                               & Real Estate                                                                   & PXR1           & \multicolumn{1}{c|}{\begin{tabular}[c]{@{}c@{}}\revised{A real estate company alters VR reconstructions of properties, so that memories of potential }\\ \revised{buyers who revisit them in VR are distorted and more positive, to influence them to buy them.}\end{tabular}}    &                                                                                                                      \\ \cline{1-4}
S16                               & Judge Training                                                                & PM1            & \multicolumn{1}{c|}{\begin{tabular}[c]{@{}c@{}}\revised{A judge training VR app proposes to visit reconstructions of previous trials. These} reconstructions\\ are altered by changing people's gender and race to help judges reducing bias.\end{tabular}}                       &                                                                                                                      \\ \cline{1-4}
S17                               & Parallel Album                                                                & PXR3           & \multicolumn{1}{c|}{\revised{An attacker alters VR family albums of an old man, }to blur their souvenirs and make them vulnerable.}                                                                                                                                               &                                                                                                                      \\ \cline{1-4}
S18                               & Buy a Friend                                                                  & PXR4           & \multicolumn{1}{c|}{\begin{tabular}[c]{@{}c@{}}A VR social media to share immersive memories allows premium users to be inserted \\ in other users' memories (fig.\ref{fig:scenario_buyFriend}).\end{tabular}}                                                                    &                                                                                                                      \\ \cline{1-4}
S19                               & Burning Stove                                                                 & PXR6           & \multicolumn{1}{c|}{\begin{tabular}[c]{@{}c@{}}\revised{An attacker displays previously experienced events in the AR user's field of view, }\\ \revised{to make them mix memories of the past with the present, and make them vulnerable.}\end{tabular}}                          &                                                                                                                      \\ \cline{1-4}
S20                               & What's Up Girl                                                                & PXR5           & \multicolumn{1}{c|}{\begin{tabular}[c]{@{}c@{}}VR is used to relieve past experiences. \revised{An attacker alters the recordings, to}\\ \revised{distort people's memories and alter their relationships.}\end{tabular}}                                                         &                                                                                                                      \\ \hline
S21                               & AR Eraser                                                                     & PXR2           & \multicolumn{1}{c|}{\begin{tabular}[c]{@{}c@{}}\revised{An AR app proposes to help forgetting someone, by hiding all the objects linked }\\ \revised{to that person (fig.{\ref{fig:scenario_arEraser}}).}\end{tabular}}                                                           & \begin{tabular}[c]{@{}c@{}}Pre-retrieval \\  Induce forgetting \\ by removing cues\end{tabular}                      \\ \hline
S22                               & False Training App                                                            & PXR1           & \multicolumn{1}{c|}{\begin{tabular}[c]{@{}c@{}}In a VR quiz application, users are immersed in a stressful environment to make them forget\\ the answers and feel incompetent, to make they buy a training product.\end{tabular}}                                                 & \begin{tabular}[c]{@{}c@{}}Pre-retrieval \\ Induce forgetting \\ by inducing stress\end{tabular}                     \\ \hline
\multicolumn{1}{l}{\revised{S23}} & \revised{College Dropout Spam}                                                & \revised{PM5}  & \multicolumn{1}{c|}{\revised{An AR app adds advertisement in the user's field of view in exchange of free education content.}}                                                                                                                                                    & \multirow{3}{*}{\revised{Not considered as XRMM}}                                                                    \\ \cline{1-4}
\multicolumn{1}{l}{\revised{S24}} & \revised{Smell}                                                               & \revised{PXR6} & \multicolumn{1}{c|}{\begin{tabular}[c]{@{}c@{}}\revised{An XR app to consult products adds smells to trigger emotion and create a comfortable}\\ \revised{environment, to make users buy their products.}\end{tabular}}                                                            &                                                                                                                      \\ \cline{1-4}
\multicolumn{1}{l}{\revised{S25}} & \revised{Stranger Things' Effect}                                             & \revised{PM5}  & \multicolumn{1}{c|}{\begin{tabular}[c]{@{}c@{}}\revised{A VR entertaining app integrates cultural content related to the user's youth (e.g., music), }\\ \revised{to make people enjoy the content.}\end{tabular}}                                                                &                                                                                                                      \\ \hline
\end{tabular}
}
\captionof{table}{\revised{Short description and categorization of the XRMM scenarios generated by participants and discussed during the workshops.}}
\label{fig:scenarios} 
\Description{First column: Id of the scenario, used to refer to it in the paper. Column 2: Name of the scenario. Column 3: ID of the participant who generated this scenario. Column 4: a few sentences describing the scenario. Column 5: class of XRMM of this scenario.}
\end{table*}

\end{document}